# HOLISTICALLY PLACING THE ICT ARTEFACT IN CAPABILITY APPROACH

Mathew Masinde Egessa, Technical University of Mombasa, egessa.mathew@gmail.com

Samuel Liyala, Jaramogi Oginga Odinga University of Science and Technology, sliyala@yahoo.com

**Abstract:** This paper proposes a framework that holistically places the Information and Communication Technology (ICT) Artefact in Capability Approach (CA). The framework harmonises the different conceptualisations of technology within CA-based frameworks in ICT4D, in order to address the inconsistencies. To illustrate the framework, while simultaneously addressing the highest thematic research gap among post-2015 ICT4D research priorities, the study collected primary data from users of Pay-As-You-Go (PAYGO) Solar Home Systems who reside in rural Kenya. Using the framework, the study revealed that the ICT-artefact can holistically be conceptualised within three of CA's concepts: under material resources as a capability input; as a new category of conversion factors (technological conversion factors); and as a component within the structural context. The study further demonstrated how the same ICT artefact could play out in the three different conceptualisations, resulting in different development outcomes for individuals. The study finally presents the implications for policy and practice.

## 1.  INTRODUCTION

Information and Communication Technology for Development (ICT4D) is an emerging and vibrant field of practice and research, that focuses on the use and design of ICTs in efforts to further (socioeconomic) development (Burrell & Toyama, 2009; Kleine & Unwin, 2009; Stillman & Linger, 2009). ICT artefacts are critical components within ICT4D studies. They include "bundles of hardware infrastructure, software applications, informational content, and supporting resources that serve specific goals and needs in personal or organizational contexts" (Mehdi, 2018, p. 631).

Despite different conceptualisations of development within ICT4D, Capability Approach (CA) has emerged as the holistic lenses to conceptualise development. Over a dozen CA-based frameworks have been developed to operationalise CA in ICT4D studies (Hatakka and Dé 2011; Kivunike et al. 2014; Kleine 2010; 2013; Zheng and Walsham 2008;). However, the ICT artefact, or technology (more broadly defined), is not conceptualised in a consistent way across the different frameworks. There is still no consensus in these frameworks, on how 'technology' relates to the core concepts of CA (inputs or 'resources', capabilities, conversion factors, functionings, structural constraints and agency) (Haenssgen & Ariana, 2018, p. 99).

This paper makes a case for a re-configured analytical framework that holistically places the ICT artefact within CA. It illustrates the framework using an ICT-enabled, renewable energy intervention, in rural Kenya.

## 2. LITERATURE REVIEW

### 2.1. Capability Approach

CA is a normative theoretical framework for the evaluation and assessment of individual well-being and social arrangements; the design of policies; and proposals about social change. Its core idea is that social arrangements should aim to expand people's capabilities (their freedom to promote or achieve valuable beings and doings).

Sen (1999, p. 75) defines **functionings** as "the various things a person may value doing or being". Functionings are the valuable states (beings) and activities (doings) that make up people's well-being, such as being safe; being calm; working; resting; having a warm friendship or taking part in political decisions. Functionings are related to resources (goods and income), but they focus on what a person is able to do or be as a result.

**Capability** refers to a person's or group's freedom to promote or achieve valuable functionings. Sen (1992, p. 40) posits that capability "represents various combinations of functionings (beings and doings) that the person can achieve. Capability is, thus, a set of vectors of functionings, reflecting the person's freedom to lead one type of life or another… to choose from possible livings".

In CA, the term **'resources'** is interpreted in a broader sense than the understanding of the term elsewhere in the Social Sciences. The focus is on material resources: either income and wealth or the consumption that these financial means (or unpaid production) generated. The resources and consumption could be conceptualised as capability inputs. They are means to the opportunities to be the person one wants to be, and to do what one has reason to value doing.

These resources do not all have the same power to generate capabilities. They depend on the individual's **'conversion factors',** as well as **structural constraints**. Conversion factors either filter, amplify or modify the input characteristics. They determine the degree to which a person can transform a resource into a functioning. All conversion factors influence (enable or inhibit) how a person can be or is free to convert the characteristics of the resource into a functioning, yet the sources of these factors may differ.

Conversion factors are often categorised into three groups: personal; social; and environmental conversion factors (Crocker & Robeyns, 2009, p. 68; Robeyns, 2005, p. 99). The three types of conversion factors enable the acknowledgement that it is not sufficient to only know the resources that a person owns, or can use, in order to be able to assess the wellbeing that he or she has achieved or could achieve; rather, there is need to know more about the person and circumstances in which he or she is living.

**Structural constraints** can have a great influence on the conversion factors as well as on the capability sets directly. There is a difference between social conversion factors and structural constraints. While structural constraints affect a person's set of conversion factors including the social conversion factors that s/he faces, conversion factors only help to convert characteristics of resources into capabilities. Structural constraints affect conversion factors but can also affect a person's capability set without impacting on the conversion of resources into capabilities (Robeyns, 2017).

Figure 1 depicts a stylised visualisation of the core concepts of CA

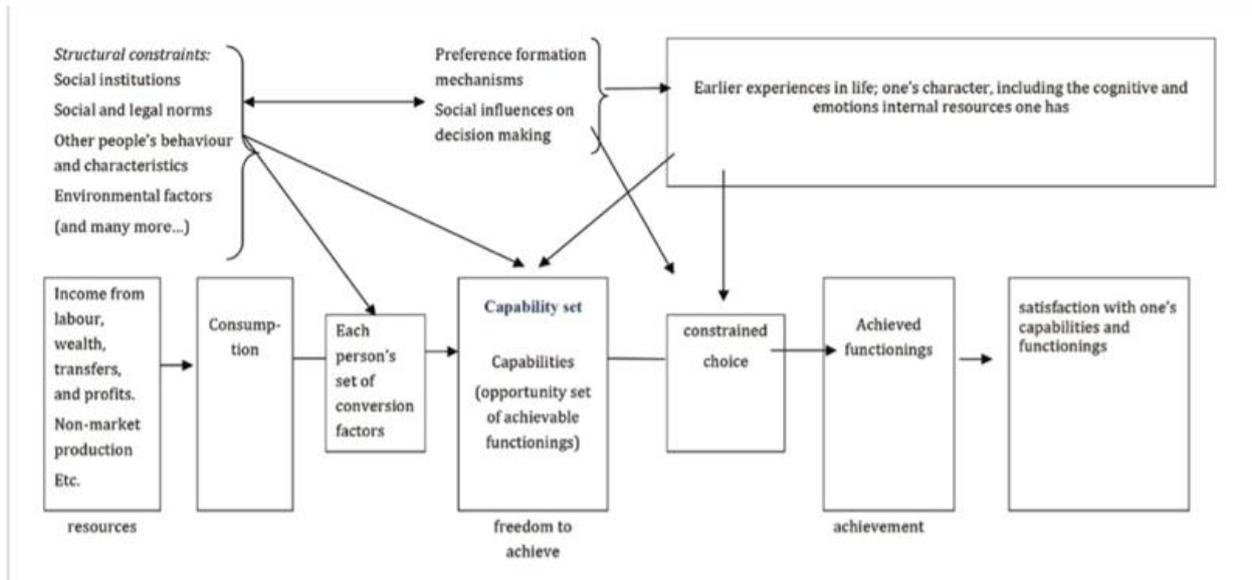

**Figure 1: A Stylised Visualisation of the Core Concepts of CA (Source: (Robeyns, 2017, p. 83))**

## 2.2. Choice Framework

The Choice Framework (CF) by Kleine (2010, 2013) is arguably the most widely used operationalisation of CA in the field of ICT4D (Sein et al., 2019; Zelezny-Green, 2018). However, it falls short in remaining consistent with this study's reading of Sen's core concepts and terminologies. Figure 2 shows the visualisation of the core concepts and terminologies within the CF.

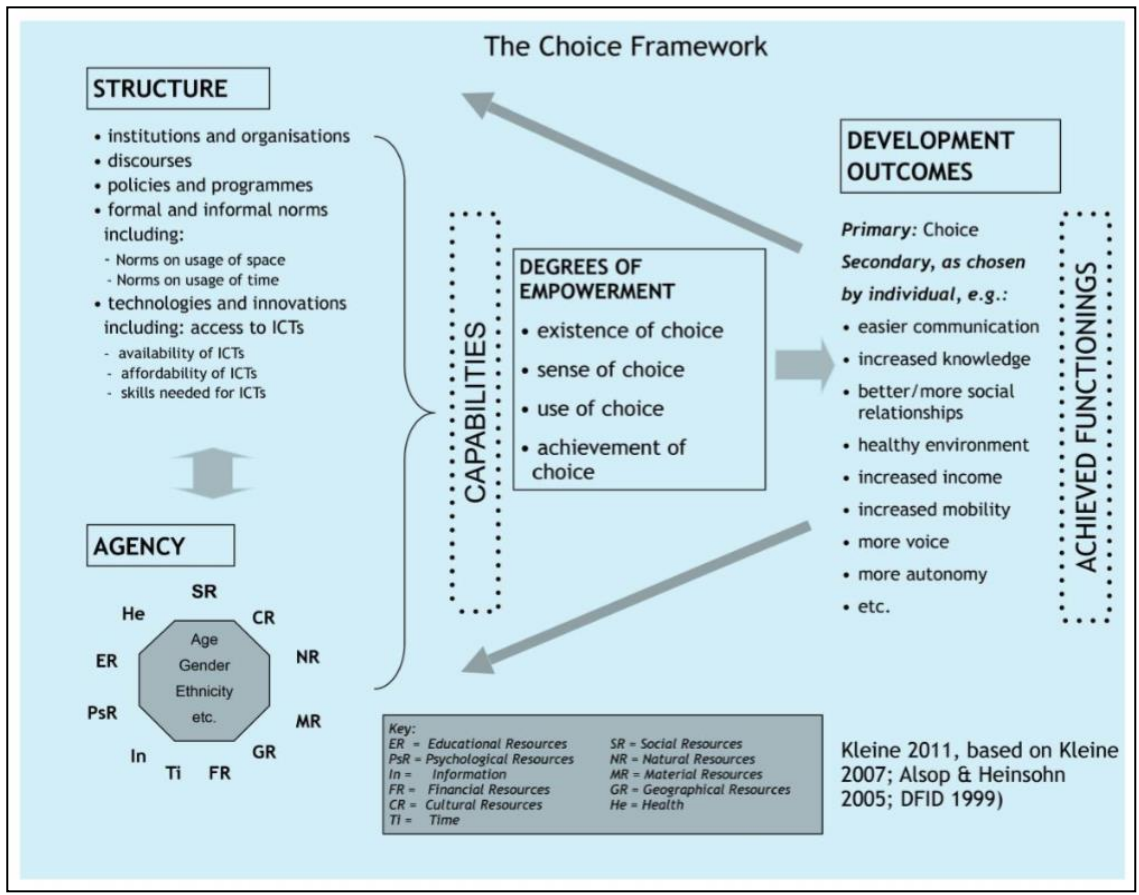

**Figure 2: Choice Framework (Source: (Kleine, 2013, p. 122))**

## 3.     PROPOSED FRAMEWORK

This paper intends to reduce the difficulty in operationalising CA within ICT4D, as well as remaining consistent with Sen's core concepts and terminologies. It proposes having a reconfiguration and realignment of the various agency-based and structure-based capability inputs of CF into conversion factors in Robeyns (2017) stylised presentation of CA. The paper then harmonises the different conceptualisations of technology within CA-based frameworks in ICT4D. The paper further proposes the explicit and holistic addition of the ICT artefact as a component within the structural context, thus extending Haenssgen & Ariana's (2018) conceptualisation.

Kleine (2010, 2013) does not use the term 'conversion factors' in her CF. However, some of the resources in her individual-based capability inputs are described in a way that fits Sen's conceptualisation of 'conversion factors'. If the CF is adopted to evaluate the development outcomes in an ICT4D intervention, the material resources of the intervention become what is 'consumed' to generate capabilities and later functionings. The material resources are the physical objects with characteristics that enhance human capabilities directly. All the other resources in the portfolio will only play a facilitative role in the conversion process, that is, if at all they come into play. At the root of the analysis, it will boil down to a material resource that will provide the main 'generative' characteristics to achieve a functioning. The material resource could still double up in other places of the CA analysis (as this paper will later demonstrate), but at the core, it will act as the capability input.

To differentiate between resources and conversion factors, working backwards from an achieved functioning, an individual would not have achieved the functioning without the presence of the resource. However, with the absence of a conversion factor, the same individual could still have achieved the functioning, but to a different degree or intensity.

In this study's reconfiguration of the agency-based capability inputs by Kleine (2013), it is only material resources that are retained as a capability input. The rest of the resources are redistributed into the three conversion factors as described by Sen. Educational, psychological, information, health and financial resources are reconceptualised as personal conversion factors. Time, social and cultural resources are reconceptualised as social conversion factors. Lastly, geographical and natural resources are reconceptualised as environmental conversion factors. The reconfiguration and pairing up of the resources to the categories of conversion factors by Sen, is achieved by matching up their corresponding definitions (ibid).

Within CF, the structures that frame people's lives are related to the concept of structural constraints in CA. They include: institutions and organisations, discourses, policies and programs, formal and informal norms, as well as technologies and innovations. Contrary to Haenssgen & Ariana's (2018, p. 101) reading of CF's structures, and equating them to conversion factors, this study closely associates CF's structure with the structural constraints in the abstract CA. CF's structure is related, but not equivalent to conversion factors. It is more equivalent to structural constraints, as Robeyns (2005, 2017) defines it, in relation to CA.

In the proposed framework, 'structural context' is conceptualised as an overarching construct that affects and gets affected by almost all the other constructs in the framework. Structural context encompasses both the structure in CF and the structural constraints in CA. However, this study refrains from using the term 'constraint' because it connotes an inhibition, yet, the same construct could as well play an enhancement role to the other concepts in the framework.

Overall, this study proposes that the ICT-artefact be holistically placed within three of the constructs in the framework: under material resources as a capability input; as a new category of conversion factors (technological factors); and as a component within the structural context.

First, this paper is in agreement with Haenssgen & Ariana's (2018) argument that, in most of the operationalisations of CA, ICTs have been conceptualised as an input that enables capabilities, 'used' alongside other resources like food, to exploit its particular characteristics. Sen (Sen, 2010, p. 2) argues that the role of the mobile phone is typically 'freedom-enhancing' so that, as a resource, it is subject to conversion factors such as computer literacy and infrastructural context. The ICTs such as mobile phones have intrinsic characteristics that can expand human capabilities and therefore fulfil the same purpose as other inputs in CA. This 'generative' dimension of the ICT-artefact qualifies it for explicit inclusion as a material resource in CA analysis.

Secondly, in other operationalisations of CA, ICTs are conceptualised in a broader way. They are understood to interact with other conversion factors and influence how inputs are used. Gigler (2004, p. 9) states that 'ICTs can play an important role not only in their own right, but can act as 'agents' for the strengthening of the poor's capitals in multiple areas'. This operationalisation of CA therefore indicates the 'transformative' dimension of ICTs. Likewise, Heeks and Molla (2009, p. 34) contend that apart from being resources, ICTs can fit as conversion factors in CA. For example, technological objects can alter the characteristics (e.g nutritional content or taste) of other inputs (e.g food) by modifying them directly (e.g through a cooking stove).

In agreement with the arguments of Haenssgen & Ariana's (2018) on the 'transformative' dimension that technological objects possess, this study further proposes the explicit addition of another factor (technological factors), to the other three conversion factors (personal, social and environmental). This is necessitated because technological objects do not perfectly fit into any of the other factors, they neither conform to personal, social nor environmental factors.

Thirdly, taking the argument forward, since this study equated structure in CF to the 'structural context' in the proposed framework, then technology will also have an explicit place in the 'structural context'. Kleine (2010, 2013) includes technology and innovations within elements of structure. Zheng (2009) argues that technology (especially ICT) co-evolves with values and choice processes.

This paper's placement of ICTs within the 'structural context' is also in line with Heeks and Molla (2009, p. 34) who also argue that ICTs can act as 'choice developers' because they can change perceptions of personal needs and preferences. Figure 3 depicts the proposed framework.

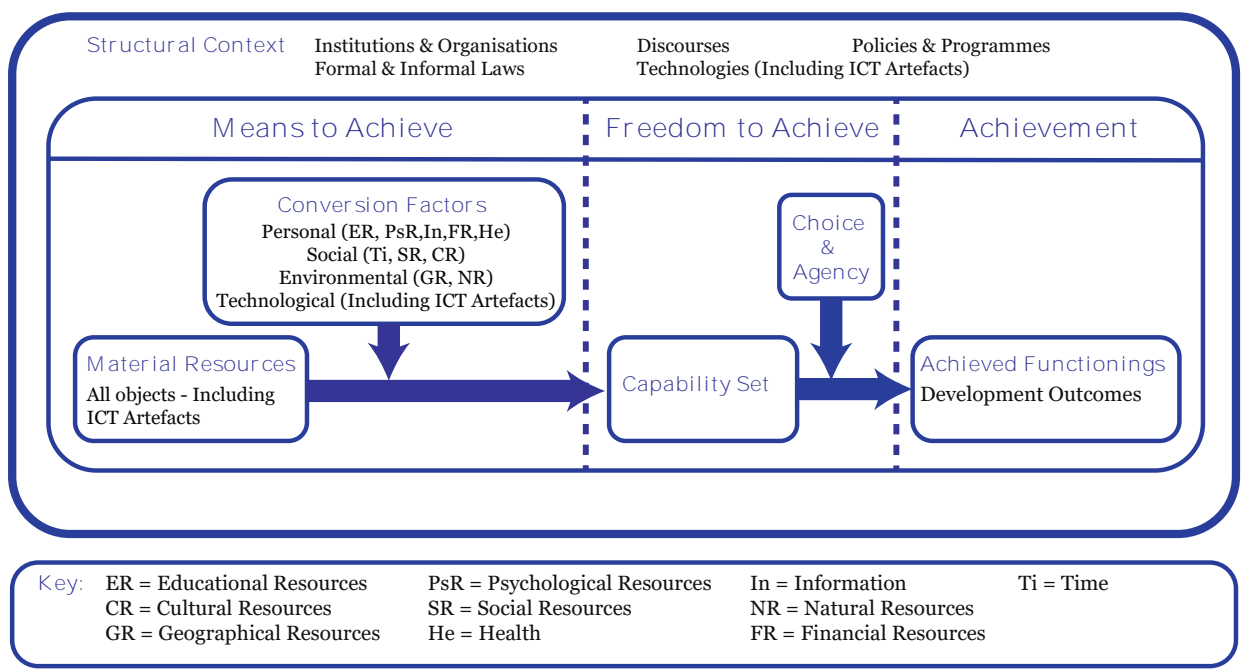

**Figure 3: Holistic Placement of the ICT Artefact in Capability Approach**

# 4. ICT-ENABLED SOLAR SOLUTIONS

Topically, 'e-Environment and Sustainable Informatics' is considered to have the highest research gap among post-2015 ICT4D research priorities (Heeks 2014). The Sustainable Development Goals (SDGs) have included a specific target (SDG 7.1) solely for ensuring access to affordable, reliable and modern energy for all by 2030. This asserts the key role that access to modern energy services, plays in achieving the other SDGs (United Nations, 2015, 2017). To demonstrate the significance of the developed analytical framework, while equally addressing this topical research gap, this study collected primary data from users of Pay-As-You-Go (PAYGO) solar kits who reside in rural Kenya.

PAYGO denotes to an assortment of technologies, ownership modes, payment arrangements and financing structures that allow a solar kit customer to pay in instalments. The embedded Machine-to-Machine (M2M) connectivity disables the solar kit if a payment is overdue (G.S.M.A., 2016; M-KOPA, 2016). The kit will not discharge power until a payment is made.

At a vendor's shop, the client normally pays around 30 USD as the initial amount, for a basic Solar Home System (SHS). The SHS consists of a battery and a control unit, a Photo Voltaic (PV) panel, a phone charger, two or three Light Emitting Diode (LED) bulbs, and sometimes other appliances. The buyer then makes regular (daily/weekly) payments, via the mobile phone, of 0.30 - 0.50 USD per day to access the services. After the client pays the full agreed amount, the system automatically switches to free use, requiring no further top ups. The customer then owns the system (BNEF & LG, 2016; G.S.M.A., 2017; M-KOPA, 2015).

# 5. RESEARCH DESIGN

The study adopted a constructivist ontology and an interpretivist epistemology. The study used the embedded single-case design because within the single-case (PAYGO solar intervention) there are sub-units (different providing companies/brands). The population for the study were the users of ICT-enabled Solar Home Systems who reside in Junju and Sokoke County Assembly Wards in Kilifi South and Ganze sub-counties of Kilifi County, at the Coast of Kenya. Purposive sampling was used to select study participants. For the purposes of both triangulation and transferability, the study used three data collection methods: in-depth semi-structured interviews, observations and document reviews. 24 in-depth semi-structured interviews were conducted with users of 4 different PAYGO solar companies and 4 in-depth semi-structured interviews were conducted with representatives of 4 PAYGO solar companies. The interviews, which lasted between 30 minutes and 1 hour, were audio recorded. The recordings were later transcribed and translated from Swahili to English.

# 6. FINDINGS AND DISCUSSION

Just like in CF, this study begins the analysis from the right of the proposed framework, coming to the left (Kleine, 2013). In line with Robeyns (2017), the study uses the term 'development outcomes', to refer to the achieved functionings, as a proxy to measure capabilities and wellbeing.

## 6.1. Development Outcomes Arising from PAYGO Solar Kits

For this study, development outcomes are conceptualised as valuable states (beings) and activities (doings) that make up a person's wellbeing, such as being safe, resting, and being calm. Arising from the in-depth semi-structured interviews, the research participants (RPs) enumerated many states and activities that they valued and had reason to value. They also elaborated on their aspirations. From the analysis of their responses, several development outcomes emerged. This study shall however narrow down to only those development outcomes that were closely linked to the ICT artefacts that are bundled within the PAYGO solar kits. They include**: 'communicating'; 'having additional study time'; 'increasing sense of security'; 'increased income'; 'making savings' and 'being entertained'.**

All the respondents valued **'communicating'** to people in a different location. This development outcome was made possible through the use of mobile phones. By implication and extension, all of them were happy with the PAYGO solar kits. The kits allowed them to keep their phones charged. Even before acquiring the solar kits, they kept spending cash regularly in order to keep their phones charged. This expenditure on charging their phones, despite their limited incomes, demonstrates the value that they had placed on communication via the mobile phones.

The respondents also valued **'being in a well-lit environment'** especially at night (RP10, RP15, RP6, RP12, RP9).

*"What I perceive as the benefit for me is the lighting. It makes my place well lit…. I do not want this place to remain dark"* –**Research Participant (RP) 10**

*"The solar kit is important because without it, there will be darkness yet we are used to bright light, just like during the day..."*- **RP 6**

Being in a well-lit environment is a development outcome that was closely linked to **'having additional study time'**. Most respondents who valued being in a well-lit environment (RP 3, RP11, RP17, RP10, RP2), also demonstrated desire to have their children or siblings, study for longer at night, courtesy of the bright lighting from the solar kits.

*"It has helped on the side of my children. Especially at night, they usually switch on the lights and put them on the table for reading. That has really helped me,* - **RP 3**

*"It even helps the children to read while at home"*-**RP 2**

The respondents (RP10, RP11, RP12) also valued **'increasing sense of security'**. This is also a development outcome that was also closely linked to being in a well-lit environment. It may be seen as a subsequent development outcome.

*"…I keep livestock, so the light has helped because I do not want this place to remain dark. The light enhances security by making this place open"* - **RP10**

Another main development outcome that emerged from the field visits is **'having increased income'** which is also closely tied to the development outcome of **'making savings'** on expenditure. Different respondents (RP2, RP3, RP9, RP22) were engaged in different economic activities. Therefore, the development outcome of **'increased income'** mattered to them because it was linked to the ICT artefacts that come together with the PAYGO solar kits. The ICT artefacts are either 'consumed' as a service, in exchange for cash, or they enhance an economic activity that generates cash for the owners.

*"I also charge other people's phones at a cost of only Kshs. 10…..I have some additional income. Before possessing the solar kits, I never used to charge mobile phones for people, which I now do. For the little cash that I get, I am grateful."* – **RP 3**

*"… we have placed here the TV and we charge people to enter to watch the TV. They normally start coming in at about 6.30pm and depart at about 11.00pm"* - **RP 22**

The respondents (RP2, RP3, RP8, RP13) also valued **'making savings'** on their expenses. This is demonstrated by the enthusiasm that they exhibited when describing the kind of savings that they made on foregoing kerosene purchases.

*"In terms of lighting the house, the solar product really helped a great deal because I stopped buying kerosene"* – **RP13**

*"whenever you have solar lamps, you do not have to spend on kerosene"* - **RP 8**

Lastly, the respondents also valued **'being entertained'** or **'being relaxed'.** A number of them would do it by engaging in sporting activities while others liked watching TV. Many of the residents in the villages would pay to enter video viewing shops to watch news and movies. This is demonstrated by **RP 15** who liked going to video viewing shops at the market centres to watch news in the evenings. **RP 21** and **RP 22** operated an establishment that used to charge people to watch

TV. Going by the number of people who flocked the establishment on a daily basis, it shows that the people actually valued getting entertained.

## 6.2. The Holistic Place of the ICT Artefact in CA

### 6.2.1. ICT Artefact as a Material Resource

As a material resource, the ICT artefact becomes the means to the opportunities of value, which an individual may want to be or do. In order to achieve the desired outcomes, the individual has to consume the characteristics of the material resource while navigating the structural context to achieve the capability set. The transformation from the resource to a capability will be enhanced or inhibited by conversion factors. The individual will then have to draw from his agency and decision-making mechanisms in order to choose a functioning from his capability set.

From the findings of this study's field visits, the following artefacts could be conceptualised as material resources – **phone charger, TV and radio.**

#### 6.2.1.1. Phone Charger as a Material Resource

The phone charger that comes bundled with the PAYGO solar kits can be conceptualised as a material resource. This conceptualisation will fit those who use it for economic purposes. Those who charge other people's phones as a service, at a cost. To the owners of the PAYGO solar kits (**RP2, RP3** and **RP9**), the accompanying phone charger acted as a material resource that led to the development outcome of '**increased income**'.

The same phone charger in the same conceptualisation as a material resource also resulted in the development outcome of '**increased savings**' for **RP1, RP3** and **RP6.**

#### 6.2.1.2. TV as a Material Resource

The TV that comes bundled together with the PAYGO solar kits can be conceptualised as a material resource, a means to achieve the different development outcomes. From the field visits, the TV artefact was able to achieve **'increased income'** for those who were enterprising and who charged a fee, so that others could watch such as **RP22**.

On the other hand, the TV artefact was also able to be conceptualised as a material resource to achieve the development outcome of **'being entertained'** or **'being educated'** by **RP12**

Through paying before watching TV at **RP 22's** establishment, and also by indications by **RP 15** and **RP6** relatives, it is implied that they were achieving the development outcome of **"getting entertained"**

The TV becomes a material resource for a person who watches TV as a means to an end. Either the owner of the PAYGO kit, a member of the household with the TV bundled with the PAYGO, or a patron who has paid to watch the TV.

#### 6.2.1.3. Radio as a Material Resource

Just like the TV, the accompanying radio in the PAYGO solar kits can also be conceptualised as a material resource. It also provides a means for **RP18** to achieve the development outcome of **'getting entertained'.** Figure 4 depicts the ICT artefact as a material resource.

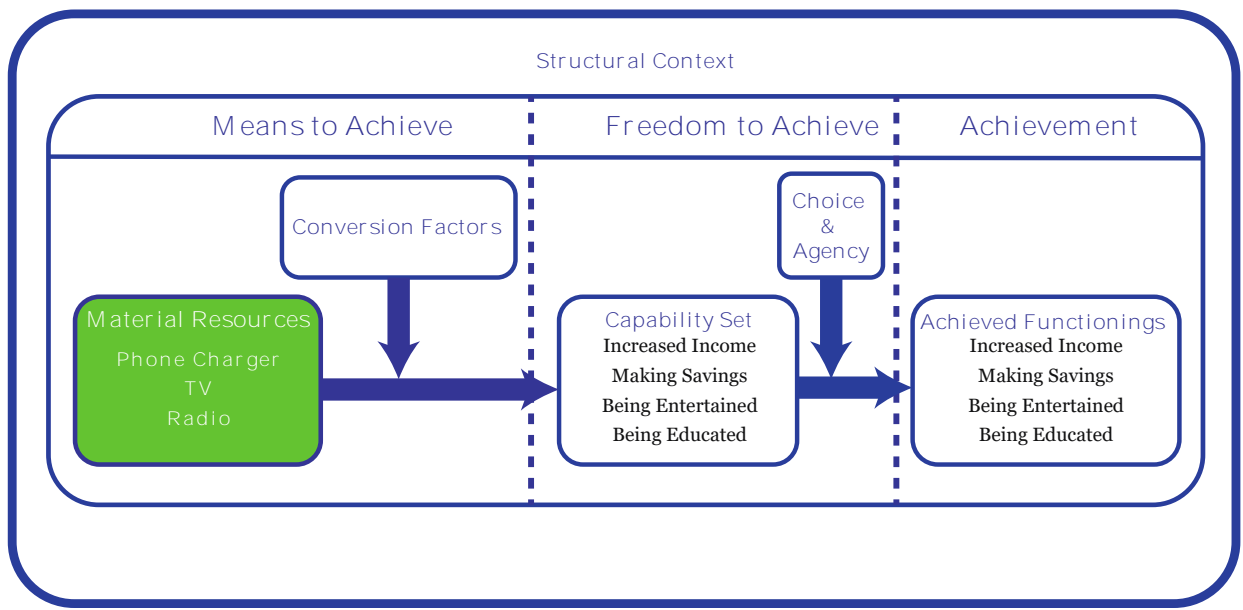

**Figure 4: ICT Artefact as a Material Resource**

This conceptualisation is in line with other scholars who have depicted in their analysis, the ICT artefact as a capability input (Haenssgen & Ariana, 2018; Ibrahim-Dasuki & Abbott, 2010; Kivunike et al., 2014; Ruhiu, 2016). They have however used different names for the ICT component: ICT intervention; commodities e.g. socio-technical interventions; ICT characteristics and technical objects.

### 6.2.2. ICT Artefact as a Conversion Factor

For this conceptualisation, the 'transformative' dimension of the ICT artefact is demonstrated. It alters the characteristics of other material resources by modifying them directly. The ICT artefact as a conversion factor either inhibits or enhances the transformation of the material resource to capabilities. The findings indicate that the phone charger, the controller & the cloud-based system (machine to machine communication device), the TV and the mobile payment service, were the ICT artefacts (within the PAYGO solar kits) that could be conceptualised as conversion factors.

### 6.2.2.1. Phone Charger as a Conversion Factor

For the phone charger to act as a conversion factor in the analysis, the end goal of the usage of the phone matters. If by using the phone, the individual has achieved a functioning, then, the phone will be conceptualised as a conversion factor.

This conceptualisation was demonstrated by those respondents who only charged their phones and those of their loved ones at no pay and for personal usage.

This conceptualisation can be contrasted with the earlier conceptualisation of the phone charger as a resource. For the previous conceptualisation, the phone charging service has to be used as an income generating venture.

### 6.2.2.2. PAYGO Controller and Mobile Payment Service as a Conversion Factor

The mechanics of how the controller works was mostly invisible to the interviewees. However, from the interviews with the providers of the PAYGO solar kits and from document reviews, it was apparent that the whole working of the PAYGO model was pegged on how the controller works. It is the core enabler and inhibitor of the usage of the PAYGO solar kits.

Though this conceptualisation seemed invisible to most respondents, it was quite apparent to **RP 15** because he encountered problems with his PAYGO solar kit. The problems were able to be sorted remotely via the controller. This enabled him to continue with the use of the other material resources provided by the PAYGO solar kit such as lighting bulbs and watching TV.

Unlike the controller as an artefact, the mobile payment as an ICT artefact was visible to the respondents. They all knew that they had to make regular payments or else, their electricity would get disconnected. They knew that it was the mobile payment that enabled their getting electricity in order for them to pursue their different development outcomes.

### 6.2.2.3. TV as a Conversion Factor

The TV artefact in the PAYGO appliances can also be conceptualised as a conversion factor. Just like the phone charger, for a TV to be the conceptualised as a conversion factor, the development outcome matters. There has to be another capability input (material resource) that gets transformed into a capability. The TV has to influence the conversion process from a resource to a capability. The presence of the TV should either increase or reduce the attainment of the development outcome.

The TV was conceptualised as a conversion factor when it was used to enhance attainment of other development outcomes. For the case of **RP 22** who owned a traditional bar, the TV was used to attract patrons. The TV used to be left on, till the patrons were done with partaking their palm wine.

Though other respondents who owned traditional bars did not get the TV with their PAYGO solar kits, they indicated similar intentions. They believed that being in possession of a TV would have attracted more customers to their bars so that they could get **'increased income'** from the sale of palm wine. Figure 5 depicts the ICT artefact as a conversion factor

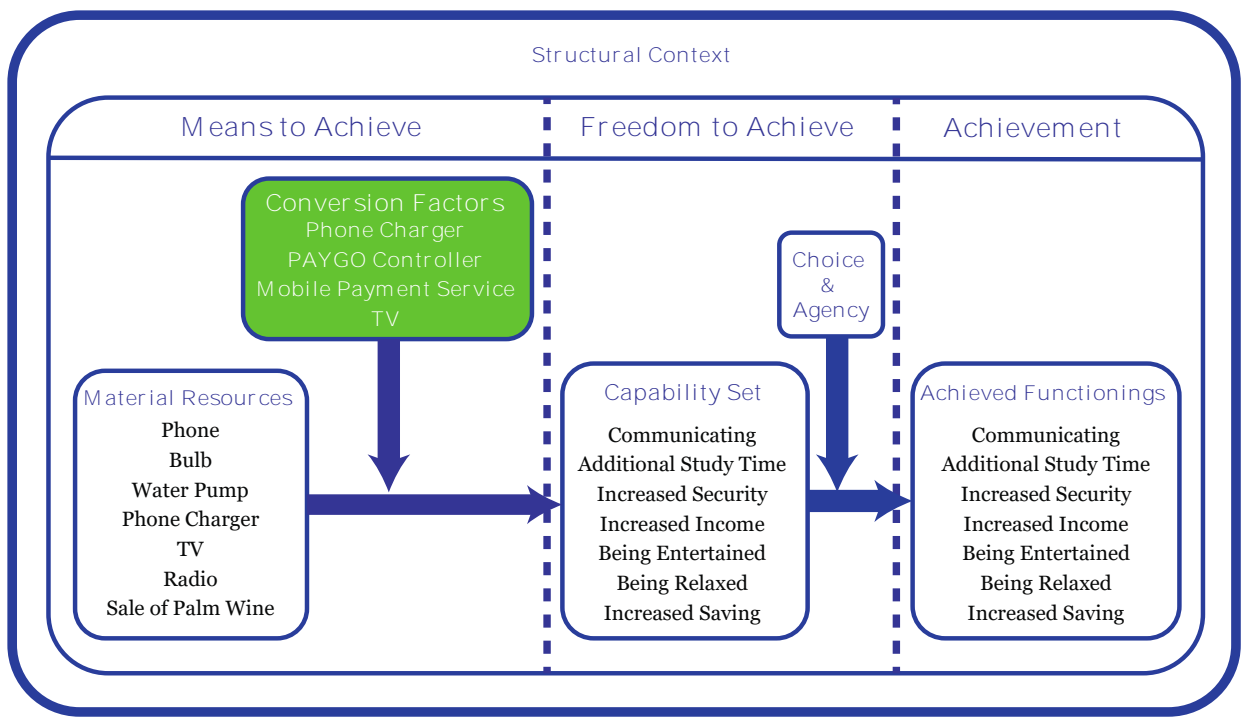

**Figure 5: ICT Artefact as a Conversion Factor**

From the proposed framework, the ICT artefact can be broadly conceptualised as a conversion factor. This is as an addition to the three traditional conversion factors (personal, social and environmental) (Haenssgen & Ariana, 2018). Conversion factors influence how other material resources are used as means to development outcomes (functionings). In this way, the ICT artefact as a conversion factor is understood to interact with other conversion factors. This conceptualisation is in line with other scholars who have alluded to such a conceptualisation but have not gone ahead to develop a framework that explicitly places it as a conversion factor (Gigler, 2004; Heeks & Molla, 2009; Zheng, 2009).

### 6.2.3. ICT Artefact as a Structural Context Component

Components in the structural context can have great influence on the conversion factors as well as on the capability sets directly. This is unlike conversion factors which only help to convert characteristics of resources into capabilities. The ICT artefact conceptualised in this way should also have the ability to influence other conversion factors in the conversion process.

The findings indicate that the phone charger, TV and PAYGO controller could be conceptualised as structural context components.

### 6.2.3.1. Phone Charger as a Structural Context Component

This is the third conceptualisation of the same ICT artefact, in the same PAYGO solar kit, but intended to achieve different development outcomes. The phone charger is conceptualised as a structural context component. This conceptualisation is used when the phone gets to be conceptualised as a conversion factor.

When the phone influences the conversion of other material resources into development outcomes, by the phone charger influencing the working of the phone, then it will ultimately be influencing a conversion factor and therefore fall within our definition of what constitutes a structural context component.

In a case where an individual uses his/her phone to conduct or facilitate economic activities, then, the phone charger will influence the usage of the phone. From the field visits, **RP1** and **RP6** depend quite a bit on their mobile phones for their livelihood. **RP1** is a middleman for selling cattle. He normally gets contacted via phone, whenever there are cattle to be sold. **RP 6** normally taps palm wine from the palm trees and sells it to the traditional bars. He mostly contacts his clients via the mobile phone, to get orders. He also uses his phone to contact buyers of bananas when they are on season.

In such cases, the phone charger can be conceptualised as a structural context component because it either enhances or inhibits the usage of the mobile phones and in these two cases, the mobile phones are acting as conversion factors in getting to the capability of **"increased income".**

### 6.2.3.2. TV as a Structural Context Component

This is similarly a third conceptualisation of the TV within the proposed framework. The TV will be conceptualised as a structural context component when it alters perceptions and changes preferences. The findings demonstrate that the TV altered behaviour and change preferences. In this conceptualisation, the TV does not come between the material resource and capabilities but influences the choice and agency that occurs between the capabilities and functionings. This demonstrates the alteration in preference formation.

Figure 6 depicts the ICT artefact as a component of the structural context.

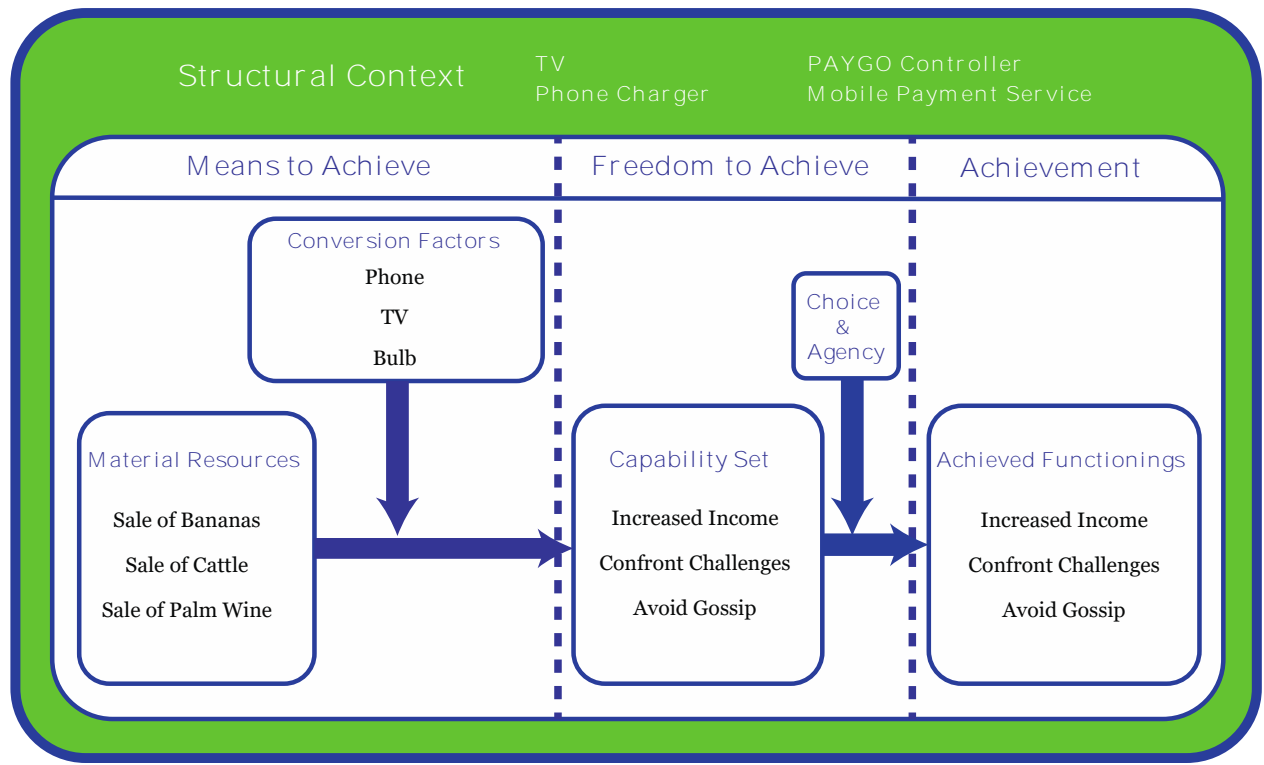

**Figure 6: ICT Artefact as a Structural Context Component**

In the proposed framework, the structural context is an overarching concept that affects and gets affected by almost all the other concepts in the framework. This is in line with Heeks and Molla's (2009) argument.

## 7. CONCLUSION AND RECOMMENDATIONS

Being attentive to the different conceptualisations of the ICT artefact as demonstrated by this study can be of help to practitioners and providers of ICT4D interventions. The providers of the ICT artefacts will be able to consider how the artefacts play out in the development journey of their customers. This will enable them to put in place mechanisms that will enhance the related constructs to help achieve the valued development outcomes.

Additionally, the different conceptualisations demonstrated by the study can be of help to policy makers, especially those seeking to create social policies that enhance people's capabilities. Having the bigger picture of all the conceptualisations of the ICT artefact will give guidance on what policies to make for greater impact.

For further research, the study recommends the use of other research designs such as multiple case studies. Employment of ethnography could also help bring out aspects about the respondents that this cross-sectional study could not.

## REFERENCES

BNEF & LG. (2016). *Off-grid solar market trends report 2016*. Bloomberg New Energy Finance and Lighting Global.

Burrell, J., & Toyama, K. (2009). What constitutes good ICTD research? *Information Technologies & International Development*, *5*(3), 82–94.

Crocker, D. A., & Robeyns, I. (2009). Capability and Agency. In C. W. E. Morris (Ed.), *Amartya Sen* (pp. 60–90). Cambridge University Press. https://doi.org/10.1017/CBO9780511800511.005

Egessa, M., Liyala, S., & Ogara, S. (2018a). Uncharted Academic Waters: A Case for mUtilities (Energy, Water and Sanitation). In R. Baguma & J. S. Pettersson (Eds.), *Proceedings of the 6th International Conference on M4D Mobile Communication Technology for Development: M4D 2018, 15-16 November 2018, Kampala, Uganda*. Karlstad University Studies. https://www.diva-portal.org/smash/get/diva2:1257356/FULLTEXT01.pdf#page=28

Egessa, M., Liyala, S., & Ogara, S. (2018b). What theory of change can contribute to capability approach: Towards evaluating ICT-enabled interventions. *2018 IST-Africa Week Conference (IST-Africa)*, Page-1.

Gigler, B.-S. (2004). *Including the excluded - Can ICTs empower poor communities? Towards an alternative evaluation framework based on the capability approach*. 4th International Conference on the Capability Approach, University of Pavia, Italy.

G.S.M.A. (2016). *Lumos: Pay-as-you-go solar in Nigeria with MTN*. http://www.gsma.com/mobilefordevelopment/wp-content/uploads/2016/11/Case-Study-Lumos-Pay-as-you-go-solar-in-Nigeria-with-MTN.pdf

G.S.M.A. (2017). *Lessons from the use of mobile in utility pay-as-you-go models*. http://www.gsma.com/mobilefordevelopment/wp-content/uploads/2017/01/Lessons-from-the-use-of-mobile-in-utility-pay-as-you-go-models.pdf

Haenssgen, M. J., & Ariana, P. (2018). The place of technology in the capability approach. *Oxford Development Studies*, *46*(1), 98–112. https://doi.org/10.1080/13600818.2017.1325456

Hatakka, M., & Dé, R. (2011). Development, capabilities and technology: An evaluative framework. *Conference Proceedings: Partners for Development- ICT. Actors and Actions*. 11th International Conference on Social Implications of Computers in Developing Countries, Kathmandu, Nepal.

Heeks, R. (2014). *Future priorities for development informatics research from the post—2015 development agenda* (Working Paper No. 57). Centre for Development Informatics Institute for Development Policy and Management, SEED. http://hummedia.manchester.ac.uk/institutes/gdi/publications/workingpapers/di/di_wp57.pdf

Heeks, R., & Molla, A. (2009). *Impact assessment of ICT-for-development projects: A compendium of approaches* (Working Paper No. 36). Centre for Development Informatics. http://hummedia.manchester.ac.uk/institutes/gdi/publications/workingpapers/di/di_wp36.pdf

Ibrahim-Dasuki, S., & Abbott, P. (2010). *Evaluating the effect of ICTs on development using the capability approach: The case of the Nigerian pre-paid electricity billing*. SIG GlobDev Third Annual Workshop, Saint Louis, USA.

Kivunike, F. N., Ekenberg, L., Danielson, M., & Tusubira, F. F. (2014). Towards an ICT4D evaluation model based on the capability approach. *International Journal on Advances in ICT for Emerging Regions*, *7*(1), 1–15.

Kleine, D. (2010). ICT4WHAT?—Using the choice framework to operationalise the capability approach to development. *Journal of International Development*, *22*(5), 674–692. https://doi.org/10.1002/jid.1719

Kleine, D. (2013). *Technologies of choice?: ICTs, development, and the capabilities approach*. MIT Press.

Kleine, D., & Unwin, T. (2009). Technological revolution, evolution and new dependencies: What's new about ICT4D? *Third World Quarterly*, *30*(5), 1045–1067. https://doi.org/10.1080/01436590902959339

Mehdi, K.-P., D. B. A. (2018). *Advanced methodologies and technologies in Artificial Intelligence, Computer Simulation, and Human-Computer Interaction*. IGI Global.

M-KOPA. (2015). *Affordable, clean energy: A pathway to new consumer choices*. http://www.m-kopa.com/wp-content/uploads/2015/10/Lightbulb-series_Paper-1-2.pdf

M-KOPA. (2016). *Digital insights build trust and enable growth*. http://www.m-kopa.com/wp-content/uploads/2016/10/1376-Digital-insights-build-trust-and-enable-growth.pdf

Robeyns, I. (2005). The Capability Approach: A theoretical survey. *Journal of Human Development*, *6*(1), 93–117. https://doi.org/10.1080/146498805200034266

Robeyns, I. (2017). *Wellbeing, freedom and social justice: The capability approach re-examined*. Open Book Publishers. https://doi.org/10.11647/OBP.0130

Ruhiu, S. N. (2016). *Exploring the conversion of ICTs to basic capabilities in community ICT interventions: Case studies in western Kenya* [PhD Thesis, University of Nairobi]. http://erepository.uonbi.ac.ke/handle/11295/97161

Sein, M. K., Thapa, D., Hatakka, M., & Sæbø, Ø. (2019). A holistic perspective on the theoretical foundations for ICT4D research. *Information Technology for Development*, *25*(1), 7–25. https://doi.org/10.1080/02681102.2018.1503589

Sen, A. (1992). *Inequality Reexamined*. Clarendon Press.

Sen, A. (1999). *Development as Freedom*. Oxford University Press.

Sen, A. (2010). The mobile and the world. *Information Technologies & International Development*, *6*(Special Edition), 1–3.

Stillman, L., & Linger, H. (2009). Community informatics and information systems: Can they be better connected? *The Information Society*, *25*(4), 255–264. https://doi.org/10.1080/01972240903028706

United Nations. (2015). *Transforming our world: The 2030 agenda for sustainable development A/RES/70/1*. United Nations. https://www.un.org/en/development/desa/population/migration/generalassembly/docs/globalcompact/A_RES_70_1_E.pdf

United Nations. (2017). *Progress Towards the Sustainable Development Goals: Report of the Secretary—General E/2017/66\**. United Nations, Economic and Social Council. https://unstats.un.org/sdgs/files/report/2017/secretary-general-sdg-report-2017--EN.pdf

Zelezny-Green, R. (2018). Mother, may I? Conceptualizing the role of personal characteristics and the influence of intermediaries on girls' after-school mobile appropriation in Nairobi. *Information Technologies & International Development (Special Section)*, *14*, 48–65.

Zheng, Y. (2009). Different spaces for e-development: What can we learn from the capability approach? *Information Technology for Development*, *15*(2), 66–82. https://doi.org/10.1002/itdj.20115

Zheng, Y., & Walsham, G. (2008). Inequality of what? Social exclusion in the e-society as capability deprivation. *Information Technology & People*, *21*(3), 222–243. https://doi.org/10.1108/09593840810896000